# A model for evolution and extinction

Bruce W. Roberts

*LASSP, Cornell University, Ithaca, NY 14853-2501*

M. E. J. Newman

*Cornell Theory Center, Cornell University, Ithaca, NY 14853-3801*

We present a model for evolution and extinction in large ecosystems. The model incorporates the effects of interactions between species and the influences of abiotic environmental factors. We study the properties of the model by approximate analytic solution and also by numerical simulation, and use it to make predictions about the distribution of extinctions and species lifetimes that we would expect to see in real ecosystems. It should be possible to test these predictions against the fossil record. The model indicates that a possible mechanism for mass extinction is the coincidence of a large coevolutionary avalanche in the ecosystem with a severe environmental disturbance.

## I. INTRODUCTION

In this paper we consider processes linking the evolution and extinction of a species. Evolution takes place when the frequency with which a gene occurs in a population of individual organisms changes. Such gene–frequency changes can arise through mutation or through selection driven either by interactions among individual organisms or by abiotic factors. This genetic viewpoint is referred to as *microevolution*. Evolution can also be described from an ecological, or *macroevolutionary* point of view by considering phenomena such as speciation, extinction, and adaptation, in which species or higher taxonomic divisions are the fundamental unit (Eldredge, 1989; Hoffman, 1989). Models for macroevolution fall into two general classes: *hierarchical* and *reductionist*. Hierarchical models represent evolution as occurring on a number of different levels, with different fundamental laws operating at each level. In contrast, the reductionist viewpoint considers evolution at all levels to be the result of the accumulation of microevolutionary changes, and the only fundamental laws are those of genetics.

We can view these two approaches to macroevolution as complementary descriptions. As an analogy, consider a pond full of water. We can think of the water in terms of its individual molecules (which corresponds to the reductionist approach to macroevolution) or in terms of properties like its density, temperature, and pressure, which are averages of properties of the individual molecules (corresponding to hierarchical macroevolutionary processes). Temperature and pressure are not relevant to describing a single molecule, but as descriptions of the entire system they are fundamental. The laws describing the changes in these quantities, which tell us, for example, how the density changes as the temperature is changed, are also fundamental at this level. We can view biological systems in a similar manner. Individual organisms or genes are equivalent to the water molecules, while phenomena operating at higher levels correspond to average properties like temperature and pressure. In this paper, we describe a model that makes use of this correspondence to study the processes linking evolution and extinction of species.

Extinction has played an important role in shaping the history of life on this planet. Of all the species that have existed since life first appeared here several billion years ago, only about one in a thousand still exists today. All the rest became extinct, typically within about ten million years of their first appearance. This has contributed greatly to the current diversity of life on the planet, with ecological niches being repopulated again and again as the species occupying these niches become extinct and are replaced by other species. This process has led to the testing of a wider range of morphological and behavioral traits than would have been possible by the slower process of phyletic transformation, where a single species gradually adapts to its environment.

The important role played by extinction leads us to the central question we will be addressing in this paper. Is extinction a fundamental part of the evolutionary process through natural selection, or is it dependent upon chance factors such as an introduced disease wiping out a single species or climate changes affecting large groups of species? This question has been addressed by a number of researchers in the field (see, for example the reviews by Raup (1991, 1986) and Maynard Smith (1989)), with arguments offered in support of both ideas. We believe that the truth, like so many truths in scientific debate, falls somewhere in between, and we have developed a model that demonstrates how the evolutionary process might interact with environmental stresses to produce results which should be visible in the fossil record, such as particular distributions of extinctions or species lifetimes (Newman & Roberts, 1995).

In his account of extinctions in terrestrial prehistory, Raup (1991) has examined the various sides of the debate on this fundamental question, asking whether species become extinct through "bad genes or bad luck"? By "bad genes" Raup means that a species became extinct because it was poorly adapted to its surroundings and had low reproductive success (Hallam, 1990a). This mechanism for extinction has recently been studied by Bak and Sneppen (1993) in a model where an initially well-adapted species can become less well-adapted if one or more of the species with which it interacts (through predation or competition, for example) evolves to a new form. This change can cause the well-adapted species to evolve itself, with the ancestral species disappearing (anagenesis or "pseudoextinction"), or to become extinct altogether. As an example, if cheetahs on the African plain get faster, then either impalas will get faster due to selective pressure or they will become extinct as they fall prey to the faster cheetahs. Such co-evolutionary arms races (Vermeij, 1987; Jackson, 1988) can lead, in Bak and Sneppen's model, to an avalanche effect in which the evolution of one species causes other species to evolve, which in turn cause yet more species to evolve, as the effect propagates through the ecosystem. They suggest that this mechanism may explain the extinctions seen in the fossil record, and that extinctions may not depend on external physical factors. A similar suggestion, which has been called the "Red Queen Model", has been proposed by Van Valen (1973; Slatkin and Maynard Smith, 1979; Stenseth and Maynard Smith, 1984). In this model, species must continually evolve to keep up with other species, in much the same way that the Red Queen in Lewis Carroll's *Through the Looking Glass* had to keep running to stay in the same place. An alternative scenario is the Stationary Model proposed by Stenseth and Maynard Smith (1984), which assumes that evolution is driven primarily by abiotic factors, and that in a constant environment evolution will cease. These two models have been tested against the fossil record by Hoffman and Kitchell (1984), who suggest that the Red Queen Model more closely describes the system they are studying. Since both models make predictions based on a constant (in time) physical environment, this conclusion is not definitive due to the difficulty of correcting the models for the varying external environments found in the Earth's geological history.

However, models such as the Red Queen Model and the model of Bak and Sneppen cannot tell the whole story since there are extinctions that are known to have an external, abiotic component (Raup & Boyajian, 1988). In fact, current paleontological evidence suggests that abiotic factors may be more important than biotic factors in controlling the organic turnover through time (Benton, 1987; Hallam, 1990b). There is also evidence that the best known mass extinction of all time, that at the K–T boundary, was influenced, at least in its later stages, by the impact of a meteor or comet about ten kilometers in diameter on the Yucatán Peninsula in Mexico (Alvarez et al., 1980; Swisher et al., 1992; Sharpton et al., 1992; Glen, 1994). (There are suggestions of other extinctions occurring shortly before the K–T boundary (Keller, 1989), so there may be more than one cause for this mass extinction. We will discuss this point further in Section IV.) These abiotic effects are what Raup called "bad luck", with species becoming extinct due to random, external factors. This is at odds with the model proposed by Bak and Sneppen, in which the causes of extinction are purely biotic. Bak and Sneppen themselves make the point that the mechanism of their model is not the only one for extinction, but that in the absence of other mechanisms, theirs might still give rise to mass extinction events.

The idea that a species might not have survived because it was better adapted than other species, but rather that it was lucky and was not affected by the particular random factors occurring at the time is known as "historical contingency". As an example, the Burgess Shale contains a fascinating diversity of organisms, many of which have left no living descendant species (Gould, 1989). The only chordate among the Burgess Shale organisms (the probable ancestor of vertebrates) was a minor species at the time, but now vertebrates are common and dominate niches filled by large animals. The Burgess organisms were themselves probably adaptive radiations which took advantage of niches left empty by an earlier, Precambrian, mass extinction event (Seilacher, 1984; McMenamin, 1990). Another example of historical contingency is the repopulation by mammals of various niches that were occupied by dinosaurs before the Cretaceous–Tertiary mass extinction. The mammals did not out-compete the dinosaurs (which had dominated terrestrial ecosystems for nearly 165 million years), but moved into niches left empty following their extinction. This idea of historical contingency plays a key role in the description of the patterns of life's history. We wish to incorporate this idea into our model for mass extinction.

One of the difficulties in modeling evolution is the problem of precise, quantitative comparison with the fossil data. Some controversies current in paleontology have not yet been settled by recourse to the fossil record, and in fact the issues might not be resolvable from the empirical evidence. These controversies include the issues of periodicity of mass extinctions (Sepkoski, 1990), whether mass and background extinctions are distinct (Jablonski, 1986), and whether species selection is an operating process (Hoffman, 1989). Other questions arise over the time scale of evolution, as in the punctuated equilibrium approach to speciation (Gould & Eldredge, 1993), and over the mechanisms for the development of species diversity.

Even though our model is a simple one, it can be difficult to test empirically. The model makes predictions that can be compared to the fossil record, but it is unlikely that this comparison will definitively accept or reject the model, at least with the present data. The problem is that the fossil record is imperfect: processes such as decay and transport—so-called taphonomic processes—can introduce bias into the fossil record by preferentially preserving certain types of organisms (Allison, 1990). For example, animals with mineralized skeletons, such as corals, are far more likely to be preserved as fossils than soft-bodied animals such as jellyfish. It is important to understand these processes because they provide us with a way of correcting this bias. Variation in the rate of deposition in sedimentary layers can also introduce bias. A mass extinction could be viewed as occurring over a short period of time if the deposition rates are very slow when it occurs, or it could be viewed as being more extended in time if taphonomic processes and subsequent erosion smear out the layer in which the extinction occurs (Signor & Lipps, 1982).

We have developed a model that includes components that represent the influence of abiotic factors. We have sought a simplified description that captures the essence of evolution, in terms of both biotic and abiotic factors, and provides insight into the processes involved. Preliminary results have been presented elsewhere (Newman & Roberts, 1995). In the present paper we present a more complete description of the theory, as well as a technique for obtaining an approximate analytic solution of the model. This technique should also prove useful for solving other dynamical models that arise in biological modeling. We also present a number of results that can be compared with paleontological data.

The paper is organized as follows. In Section II we motivate and define our model. Section III contains a description of an approximate analytic solution of the model. The results from this solution and from computer simulations are discussed together in Section IV. Conclusions are given in Section V.

## II. OUR MODEL

We have proposed a new model (Newman & Roberts, 1995) which is an extension of the model proposed by Bak and Sneppen (1993). (Further analysis of their model is given by Sneppen et al. (1995), Flyvbjerg et al. (1993), Ray & Jan (1994), de Boer et al. (1994, 1995), Paczuski et al. (1994), Marsili (1994), Maslov et al. (1994), and Jovanović et al. (1994).) Our model combines "bad genes" and "bad luck" to make predictions about extinctions and species lifetimes, and also examines the interplay between the two. The environmental influences that we include represent a form of the "historical contingency" discussed in Section I.

The model is based on the idea that species undergo evolution (or coevolution) in bursts. During these bursts, species are less well adapted to their environment and are susceptible to external stresses. Thus, the extinction rate will be higher during coevolutionary avalanches for the same level of stress than it would be during times of relative phenotypic stability. This idea has been suggested by a number of other authors (Quinn & Signor, 1989; Kauffman, 1993; Kauffman & Johnsen, 1991; Plotnick & McKinney, 1993; Parsons, 1993, 1994a, 1994b). Our simple model shows features seen in the fossil record. An example is the bursts of evolutionary activity which have been likened by Sneppen and coworkers (1995) to the punctuated equilibria postulated for the speciation rates in individual species by Eldredge and Gould (1972; Gould & Eldredge, 1977, 1993). We also see a power law distribution of extinction sizes, with extinctions ranging from ones that wipe out a large fraction of existing species to small ones that affect only a few species. We see precursor extinctions, where a series of smaller extinctions precedes a major extinction event, and aftershocks in which opportunistic but not particularly well-adapted species quickly become extinct as they rise up in the aftermath of a major event.

In evolutionary biology there are typically three time scales of interest (Stenseth and Maynard Smith, 1984): 1. the time scale of changes in the population of a species, 2. the time scale of changes in gene frequency (the proportion of individuals in a population that have a particular gene) in a population, and 3. the time scale for extinction and speciation of species. Our model will focus on a time scale intermediate between 2. and 3. We use such a time scale because we want to describe what happens when we incorporate gene frequency changes in some coarse-grained, long-time manner, with external influences that can cause extinction and speciation.

There are a fixed number $N$ of species in our model. Each species interacts with $K - 1$ other species. (We choose this notation to be compatible with the $NK$ models of Kauffman and others (Kauffman, 1993; Kauffman & Johnsen, 1991; Flyvbjerg & Lautrup, 1992; Bak et al., 1992).) Also, species are susceptible to external factors characterized by a single noise strength $\sigma$. When $\sigma \to 0$ our model reduces to the model of Bak and Sneppen. For large $\sigma$, species are wiped out at random without regard at all to genetic factors. For intermediate $\sigma$, we have the interesting regime of interaction of biotic evolution with external factors in the physical environment.

Each of the $N$ species in our model ecosystem is characterized by two real numbers, a fitness $f_i$ and a barrier to mutation $b_i$. The fitness is a measure of how susceptible a species is to extinction from environmental effects such as a climate change. Because there is no absolute fitness scale, we choose to have the fitnesses in our model lie between zero and one. The fitness is not a gauge of the relative merits of one species over another in direct competition. No mechanism for direct inter-species competition is included in the model. We could perhaps do this with interactions that can vary over time, but the resulting model is much harder to analyze theoretically and computationally. (See, for example, the work by Kauffman on $NK$ models (Kauffman, 1993; Kauffman & Johnsen, 1991).) In fact, evidence for competitive replacement playing a role in evolution is not strong (Benton, 1987).

The barrier $b_i$ to mutation is a measure of how far the species must mutate against a selection gradient (Caswell, 1989) before reaching a new evolutionarily stable phenotype. The situation is illustrated in Figure 1, which shows a portion of a "rugged fitness landscape" (Kauffman, 1993; Wright, 1982), in which the horizontal axis represents different phenotypes (or genotypes) and the vertical axis represents some measure of species success, such as average lifetime reproductive success. The landscape pictured is one-dimensional; in real biological systems it will be multi-dimensional. A species spends most of its time at a maximum in the fitness landscape, where it is reasonably well adapted to its environment. Small mutations of the species are normally driven back to the maximum by the selection gradient. On rare occasions, a species can undergo a large mutation, or a rapid succession of smaller mutations, which cause it to pass a barrier (a region of relatively low fitness) and reach the domain of attraction of a different maximum in the fitness landscape. The selection gradient will then drive it to this new maximum, where it will remain, undergoing small fluctuations about its new phenotype, until another large change drives it to a different maximum again.

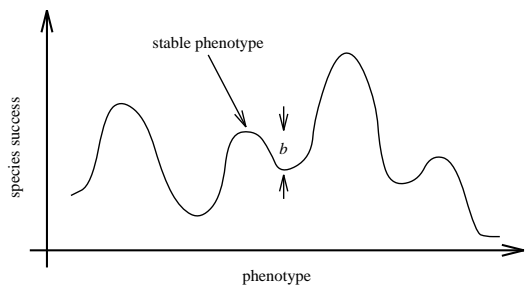

FIG. 1. A schematic representation of a portion of a rugged fitness landscape. Maxima in the fitness function correspond to stable phenotypes. The smallest fitness barrier that has to be traversed by a species at the indicated point in order for it to mutate to a new stable form is given by $b$.

Our barrier variables are a primitive representation of the situation shown in Figure 1. We take into account only the smallest barrier a species needs to overcome to reach a new stable phenotype (for the landscape shown in Figure 1, the smallest barrier is $b$). There are no obvious units for the barrier heights, so again following Bak and Sneppen (1993), we choose them to lie between zero and one. For initial conditions, we assign each of the $N$ species a barrier and a fitness chosen at random from the allowed range.

The simulation consists of the repetition, always in the same order, of three basic steps. The species with the lowest barrier to mutation (call this species $m$) evolves first, passing to a new stable phenotype. The first step is to find this species and have it evolve to some new form characterized by a new value of the fitness $f_m$ and a new barrier to mutation $b_m$, which are again chosen at random from within the allowed range. This process removes the species with the lowest barriers from the population.

The second step is to select new random values for the fitness and barrier of $K - 1$ 'neighbors' of species $m$. By neighbors we mean other species that interact with the original species in some fundamental manner, such as through predator-prey relationships, competition, or some other interaction in a food web or food chain. We view the fitness landscapes of the different species as being coupled due to these interactions, with the change in adaptive maximum of species $m$ causing a change in the fitness landscape of its neighboring species. This change is represented by choosing new random barriers and fitnesses for the neighbors. Other, more complicated (and perhaps more realistic) methods of changing the neighbors' fitnesses can also be used, although the predictions of the model do not seem to be sensitive to the exact details. After a species makes an adaptive move, it is quite likely that the next species to evolve will be one of its neighbors, giving rise to coevolutionary avalanches. A species with a high barrier, unable to mutate on its own, might therefore eventually have its barrier reduced by the mutation of a neighbor. We also note that the neighboring species can be chosen randomly, or by some spatial criterion (such as nearest neighbors in a two dimensional lattice). Most of our results are for the case where we choose the neighbors of species randomly at each time step. We note that one time step in this process represents a variable amount of time in the real system. The time to mutate over a barrier varies with the size of the barrier. We choose to use as our increment of time

$$\delta\tau = \tau_0 e^{b_m/b_0}, \qquad (1)$$

where $\tau_0$ is some base time scale, set by microevolutionary genetic interactions, and $b_0$ is a scale for the barrier heights. We can also define a version of this model (the so-called thermal version) where we don't necessarily pick the lowest barrier, but we pick in some probabilistic manner among all the barriers, with the lowest ones having the greatest probability of being chosen. However, such a method for picking barriers would greatly increase the computer resources necessary to perform the simulations.

The third and final step in our simulation models the effect of external environmental stresses on the ecosystem. We imagine that the environmental forces put some stress on the system, which will cause some species to become extinct. Usually this stress will not be severe and few species will be affected. Occasionally, a large event will occur (such as a climate change) which will affect a far greater number of species, causing a mass extinction (Benton, 1987; Parsons, 1993).

To model such processes, we choose a random number $r$ between zero and one at each time step in the simulation. This number represents the stress placed on the system at that time. All species with fitness less than this number become extinct. Their ecological niches are repopulated by newly appearing species, which in the model have randomly chosen barriers and fitnesses. We can also model deterministic effects of the physical environment by choosing $r$ in some specified manner.

We have tried a number of different forms for the random numbers (or 'noise') in the model by choosing the numbers $r$ in a variety of ways, usually with smaller values of $r$ more likely than larger values. We have used white noise with Gaussian, exponential, or bimodal probability distributions, as well as $1/f$ noise. White noise is completely uncorrelated from time step to time step. Gaussian white noise is a particular form of this, with a certain probability distribution for the random numbers (we are using $r \geq 0$), given by

$$P(r) = \frac{2}{\sqrt{2\pi}\sigma} e^{-r^2/2\sigma^2}, \qquad (2)$$

where $\sigma$ is the standard deviation that describes the strength of the noise. Exponentially distributed noise is similar to Gaussian noise, except that the probability distribution is now described by an exponential, $e^{-r/\sigma}$, instead of a Gaussian. Bimodal random numbers are chosen from two possible numbers, with a probability that can vary. In contrast to all of these, $1/f$ noise is noise that is correlated in a particular manner from time step to time step. This is represented by the Fourier transform of the time series $r(t)$ for the noise being given by

$$|\widetilde{r}(f)|^2 \propto 1/f \qquad (3)$$

for large $f$, where $f$ represents the frequency in the Fourier transformed variable $\widetilde{r}(f)$. We found that the most important predictions of the model are independent of the form of the noise we choose. It is therefore not necessary to know the exact nature of external stresses on the ecosystem in order for the model to make predictions about extinctions.

Most of our results are for uncorrelated Gaussian noise with a small standard deviation of $\sigma = 0.1$ or $\sigma = 0.2$. In the limit of $\sigma = 0$, the effects of the external environment vanish, and the fitness parameters no longer play a role in the model, since no species ever has a low enough fitness to get wiped out. In this limit, our model reduces to that of Bak and Sneppen (1993).

Our simulations consist of repeating the above processes—evolution of the species with the lowest barrier, change of the fitnesses and barriers of its neighbors, and extinction of the species with the lowest fitnesses—many times (typically $1000N$ time steps). To examine properties of the model that don't depend upon particular initial conditions, we first evolve the system to a statistically stationary state, so that our results only depend on time and space differences, and not on absolute time values, for example. Then we examine the resulting patterns of extinction, distribution of extinctions sizes, distributions of barriers and fitnesses, and distribution of species lifetimes.

Figure 2 is a preview of our results. This figure shows a time series of extinctions occurring in the model. We will discuss the distribution of the extinction sizes, as well as correlations between the extinctions, in Section IV.

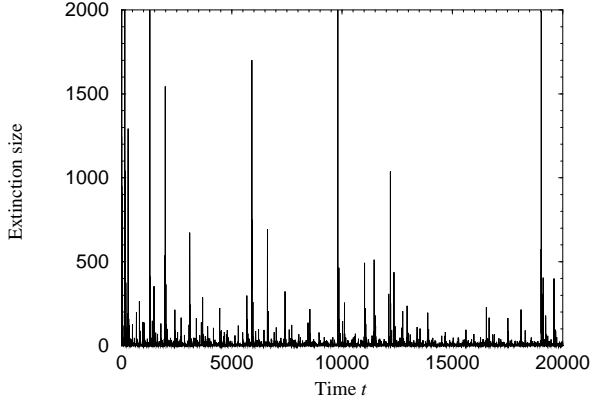

FIG. 2. A section of the extinction data from a simulation of our model with $N = 10\,000$, $K = 4$, and $\sigma = 0.1$. Notice the punctuated behavior of the model, with long periods of inactivity separated by brief bursts of heavy extinction.

## III. ANALYTIC SOLUTION OF THE MODEL

We have found an approximate analytic solution of our model. This solution is a generalization of the work of Flyvbjerg et al. (1993), who gave an approximate solution for the model of Bak and Sneppen (1993). (This work has been extended by de Boer et al. (1995).) The details of our solution are rather technical, and readers interested mainly in the results from the model may wish to skip directly to Section IV.

Here we give the solution for the random–neighbor version of the model (see Section II), in which we select the $K - 1$ neighbors of a species randomly from the $N - 1$ possibilities at each step. We let $p$ be the joint probability distribution of barriers $b$ and fitnesses $f$ at time $t$ for the entire population. Then $p(b, f, t) \mathrm{d}b \, \mathrm{d}f$ is the probability that a species has a barrier between $b$ and $b + \mathrm{d}b$ and fitness between $f$ and $f + \mathrm{d}f$ at time $t$. In order to calculate $p(b, f, t)$ we also need to know the probability distribution $p_1$ for the lowest barrier in the ecosystem, which we define such that $p_1(b, f, t) \mathrm{d}b \, \mathrm{d}f$ is the probability that the species with the smallest barrier at time $t$ has a barrier lying in the interval $b$ to $b + \mathrm{d}b$ and fitness between $f$ and $f + \mathrm{d}f$. The two distributions are related by

$$p_1(b, f, t) = N\, p(b, f, t)\, R^{N-1}(b, t) \qquad (4)$$

where

$$R(b, t) = \int_b^1 \mathrm{d}b' \int_0^1 \mathrm{d}f'\, p(b', f', t). \qquad (5)$$

This result comes from noting that the lowest barrier can be any of the $N$ species, and the probability that a particular species has barrier $b$ and fitness $f$ is given by $p(b, f, t)$; in order for this to be the lowest barrier, all $N - 1$ other species must have a higher barrier. This is represented by the factor $R^{N-1}(b, t)$, where $R(b, t)$ is the probability for one species to have a barrier higher than $b$. Noting that

$$\frac{\partial R(b, t)}{\partial b} = - \int_0^1 \mathrm{d}f'\, p(b, f', t). \qquad (6)$$

we can show $p_1(b, f, t)$ is a proper normalized probability distribution:

$$\int_0^1 \mathrm{d}b \int_0^1 \mathrm{d}f\, p_1(b, f, t)$$
$$= - \int_0^1 \mathrm{d}b\, N\, R^{N-1}(b, t) \frac{\partial R(b, t)}{\partial b}$$
$$= R^N(0, t) - R^N(1, t) = 1. \qquad (7)$$

Using equation (4), we can write down the following equations for the time evolution of the joint probability distribution $p(b, f, t)$:

$$p(b, f, t + 1/2) = p(b, f, t) - \frac{p_1(b, f, t)}{N}$$
$$- \left( \frac{K - 1}{N - 1} \right) \left( p(b, f, t) - \frac{p_1(b, f, t)}{N} \right) + \frac{K}{N}, \quad (8)$$

$$p(b, f, t + 1) = p(b, f, t + 1/2)$$
$$- \Theta[r(t) - f]\, p(b, f, t + 1/2)$$
$$+ \int_0^{r(t)} \mathrm{d}f \int_0^1 \mathrm{d}b\, p(b, f, t + 1/2), \quad (9)$$

where the function $\Theta(x)$ is a step function (it is 0 if $x < 0$ and 1 if $x \geq 0$). Equation (8) represents the "evolution" step in the model, in which the species with the lowest barrier is removed and it and its neighbors are replaced by new species. The second term on the right–hand side represents the removal of the species with the smallest barrier from the distribution. The third term represents the removal of the $K - 1$ neighbors. (These neighbors can be any of the $N - 1$ species remaining, and hence have barriers and fitnesses distributed as $[p(b, f, t) - p_1(b, f, t)/N]$.) The fourth term represents the addition of $K$ new species with evenly distributed barrier and fitness values, replacing the species removed in the second and third terms.

Equation (9) represents the "extinction" step, in which all species with sufficiently low fitness become extinct and are replaced by new species. The second term represents the removal of all the species with fitness below the randomly–chosen level $r(t)$. The third term represents the addition of species with evenly distributed barriers and fitnesses to replace those removed by the extinction step. Both equations (8) and (9) conserve probability, which can be shown by integrating them over the allowed ranges of the variables $b$ and $f$.

We wish to average these equations over the fluctuations due to the external noise. Let $D(r)$ be the probability distribution for the random noise, so that any average of a function $g$ that depends on $r$ is given by

$$\bar{g} = \int_0^1 \mathrm{d}r\, g(r)\, D(r). \tag{10}$$

Applying this to equations (8) and (9) yields

$$\bar{p}(b, f, t + 1/2) = \bar{p}(b, f, t) - \frac{\bar{p}_1(b, f, t)}{N}$$
$$- \left(\frac{K-1}{N-1}\right)\left(\bar{p}(b, f, t) - \frac{\bar{p}_1(b, f, t)}{N}\right) + \frac{K}{N}, \tag{11}$$

$$\bar{p}(b, f, t+1) = \bar{p}(b, f, t + 1/2)$$
$$- \int_f^1 \mathrm{d}r D(r)\, p(b, f, t + 1/2)$$
$$+ \int_0^1 \mathrm{d}f \int_0^1 \mathrm{d}b \int_f^1 \mathrm{d}r D(r)\, p(b, f, t + 1/2), \tag{12}$$

where in the last term we have interchanged the order of integration, with the necessary changes to the limits of integration. We now make an approximation by replacing the functions $p(b, f, t)$ and $p(b, f, t+1/2)$ occurring in these equations with their average values $\bar{p}(b, f, t)$ and $\bar{p}(b, f, t + 1/2)$. This approximation neglects correlations in the distributions from one time step to the next, and is similar in spirit to the mean–field approximation used in statistical physics (Plischke & Bergersen, 1989).

We can solve equations (11) and (12) by iterating them from given initial conditions, and reach a steady–state solution much faster than we can obtain results by performing the full simulation. We can also look for a steady–state solution to these equations in which $\bar{p}(b, f, t+1) = \bar{p}(b, f, t) = \bar{p}$ and $\bar{p}_1(b, f, t+1) = \bar{p}_1(b, f, t) = \bar{p}_1$. This solution satisfies

$$0 = -\frac{1}{N}\bar{p}_1 - \frac{K-1}{N-1}\left[\bar{p} - \frac{1}{N}\bar{p}_1\right] + \frac{K}{N}$$
$$-E(f)\left(\bar{p} - \frac{1}{N}\bar{p}_1 - \frac{K-1}{N-1}\left[\bar{p} - \frac{1}{N}\bar{p}_1\right] + \frac{K}{N}\right)$$
$$+ \int_0^1 \mathrm{d}b \int_0^1 \mathrm{d}f E(f)\left(\bar{p} - \frac{1}{N}\bar{p}_1\right.$$
$$\left. - \frac{K-1}{N-1}\left[\bar{p} - \frac{1}{N}\bar{p}_1\right] + \frac{K}{N}\right) \tag{13}$$

where

$$E(f) \equiv \int_f^1 \mathrm{d}r\, D(r), \tag{14}$$

which is the average probability that a species with fitness $f$ will survive to the next time step.

We can obtain an approximate solution to equation (13) in the limit of large $N$ if we note that $R(b) \leq 1$ with equality occurring only when $b = 0$. Then for large $N$, $R^{N-1}$ is small, and the terms in which it appears can be ignored. This leaves the equation

$$\bar{p}(b, f)\left[\frac{K-1}{N-1} + \frac{N-K}{N-1}E(f)\right] = -\frac{K}{N}E(f) + A \tag{15}$$

where the constant $A$ represents terms that have no functional dependence on $b$ or $f$. This equation has the solution

$$\bar{p}(b, f) = \frac{-\frac{K}{N}E(f) + A}{\left[\frac{K-1}{N-1} + \frac{N-K}{N-1}E(f)\right]} \tag{16}$$

where $A$ is given by the consistency condition that the probability integrated over all allowed barriers and fitnesses be 1:

$$A = \left[1 + \int_0^1 \mathrm{d}f \frac{\frac{K}{N}E(f)}{\left[\frac{K-1}{N-1} + \frac{N-K}{N-1}E(f)\right]}\right] \Big/$$
$$\left[\int_0^1 \frac{\mathrm{d}f}{\left[\frac{K-1}{N-1} + \frac{N-K}{N-1}E(f)\right]}\right]. \tag{17}$$

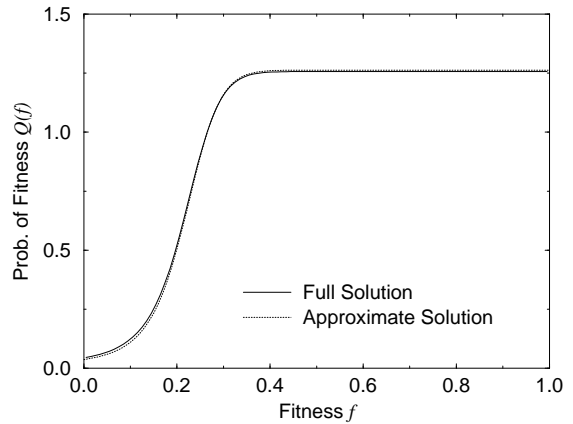

FIG. 3. Comparison of the full analytic solution to equations (11) and (12) with the approximate solution given by equations (16) and (17) for $N = 100$, $K = 4$, and $\sigma = 0.1$.

Figure 3 shows the results for the distribution of fitnesses

$$Q(f) = \int_0^1 \mathrm{d}b\, \bar{p}(b, f) \qquad (18)$$

from both an iterative solution to equations (11) and (12), in which the terms containing $R^{N-1}$ are retained, and from the approximate solution, equation (16). The agreement is excellent, and gets even better with increasing $N$. Note however that important information about the barrier height distribution enters through the term in $R^{N-1}$, so the full solution to equations (11) and (12) is still needed to obtain information about the barriers.

In addition to information about the probability distribution for the barriers and fitnesses, we can obtain information about the probability distribution for extinction sizes from the analytic solution theory. With each time step in the model, all species with fitnesses $f$ below some random number $r$ are removed. This constitutes a fraction

$$s \equiv F(r) = \int_0^r \mathrm{d}f\, Q(f) \qquad (19)$$

of the total number of species in the system. We also refer to $s$ as the extinction size. We can convert from $s$ to $r$ using

$$r = F^{-1}(s). \qquad (20)$$

We wish to calculate the probability $P(s)$ of getting an extinction of size $s$. To do this we convert from the probability distribution for the noise $r$ to that for $s$ using

$$P(s)\, \mathrm{d}s = D(r)\, \mathrm{d}r \qquad (21)$$

and hence

$$P(s) = D(r)\, \frac{\mathrm{d}r}{\mathrm{d}s}. \qquad (22)$$

From equation (19)

$$\frac{\mathrm{d}r}{\mathrm{d}s} = \frac{1}{Q(r)}. \qquad (23)$$

So we finally arrive at (using equation (20))

$$P(s) = \frac{D[F^{-1}(s)]}{Q[F^{-1}(s)]} \qquad (24)$$

So given $D$ and obtaining $Q$ and thus also $F$ from the solution to equations (11) and (12), we can determine the extinction size distribution. Some results from this will be presented in the next section.

## IV. RESULTS AND DISCUSSION

We have performed simulations of our model for systems with up to $N = 10\,000$ species and for up to ten million time steps. The length of geological time to which our time step corresponds has not been specified, but none of our results depend on knowing this. Our simple model with each species interacting with $K - 1$ others is only a crude approximation to a real ecosystem; real species can interact with other species to varying degrees, and will also interact with different numbers of neighbors. Neither of these effects is considered in the simplest version of our model, although they could be added. As a first approximation for the number of neighbors we choose $K = 4$ (as suggested by the work on food webs by Sugihara *et al.* (1989)), although the essential results are independent of the exact value of $K$ (for moderate values of $K$).

In Figure 4 we show the distribution of fitnesses (as defined by equation (18)) for two different strengths of the external noise. In Figure 5 we show the distribution of barriers

$$Q(b) = \int_0^1 \mathrm{d}f\, \bar{p}(b, f) \qquad (25)$$

for the same two cases. The symbols are the results from simulations, while the lines are values calculated from the analytic solution discussed above in Section III. The solution agrees quite well, although the agreement is somewhat better for the fitnesses than it is for the barriers. This agreement allows us to probe the behavior of our model in regimes where the statistics from the simulations are poor, and also to provide analytic insight into the nature of the extinction size distribution (equation (24)).

The behavior of the distribution of fitnesses is essentially what we would expect. The external stresses placed on the system remove the species with the lowest fitness from the ecosystem, so that the distribution has most of its weight at higher values of fitness. Near the top of the range, where species are sufficiently fit that stresses large enough to wipe them out are rare, there is no practical difference between species with different fitnesses, and the distribution is flat. In the limit $f \to 0$, the distribution goes to zero because there is always some small noise at each time step in the model that will wipe out the very lowest lying species.

The barrier distribution also tends to zero as $b \to 0$, since an infinitesimal barrier is very likely to be the lowest barrier in the ecosystem, making it very likely that it will be the next to evolve and that it will evolve into a species with a higher barrier to mutation. In the limit of large barriers, the distribution is again flat, since a species with a large barrier is unlikely to have the lowest barrier in the ecosystem, and hence is unlikely to mutate. There is then no practical distinction between the species at the high end of the distribution.

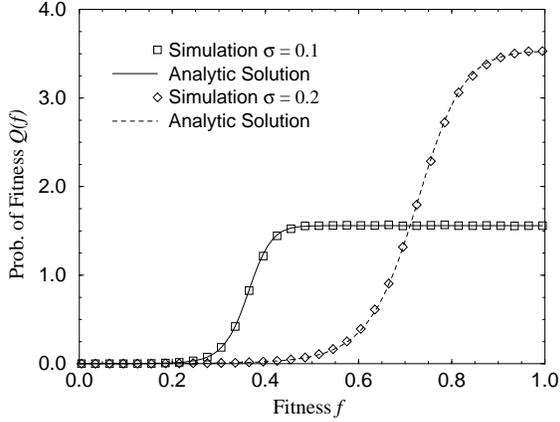

FIG. 4. A histogram of the mean fitness distribution over the whole ecosystem for two different strengths of Gaussian noise. The symbols are the results from the numerical simulations, and the solid lines are the analytic solution discussed in Section III. The parameters of the simulation were $N = 10\,000$ and $K = 4$.

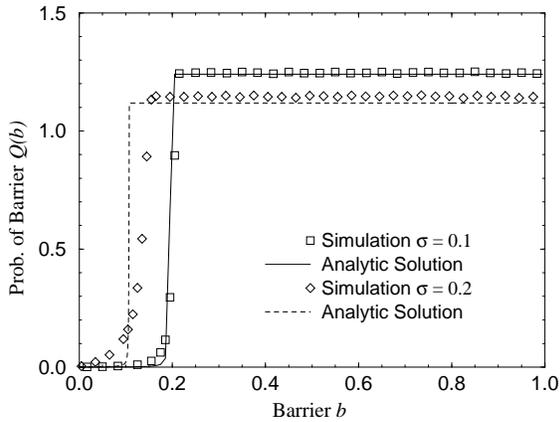

FIG. 5. A histogram of the mean distribution of barriers for two different strengths of Gaussian noise. The symbols are the results from the numerical simulations, and the solid lines are the analytic solution discussed in Section III. The parameters of the simulation were $N = 10\,000$ and $K = 4$.

In Figures 6 and 7 we show the probability distributions for barrier heights and fitnesses calculated from the analytic solution (equations (11) and (12)) for increasing $N$ with $K = 4$ and $\sigma = 0.1$ held fixed. The number of species in the terrestrial ecosystem has been variously estimated at somewhere between ten and one hundred million (of which perhaps less than one percent have been cataloged). Because of this, we then expect that the distributions for large $N$ are most relevant to results in the Earth's biosphere. For smaller values of $N$, the model could describe isolated ecosystems like islands.

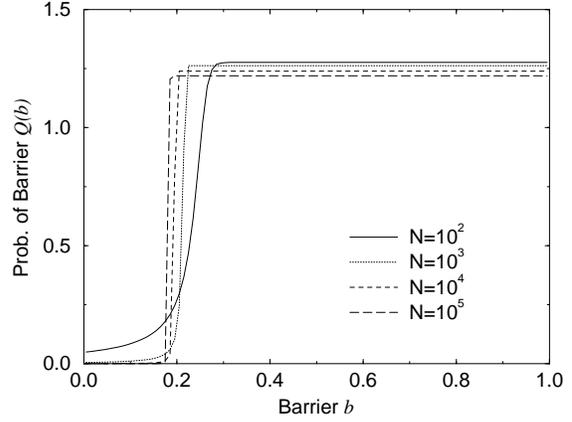

FIG. 6. Analytic result for the barrier distribution with increasing $N$ and fixed $K = 4$. The noise is Gaussian with $\sigma = 0.1$.

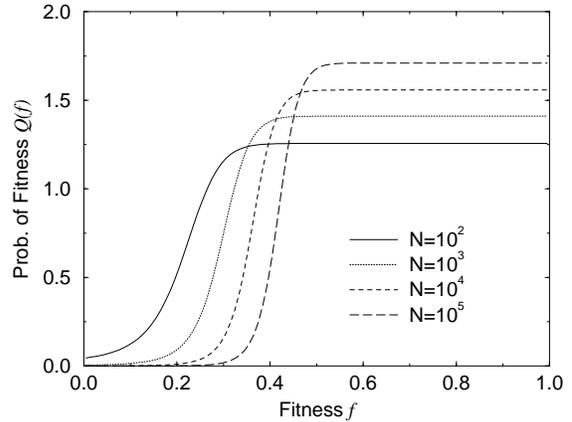

FIG. 7. Analytic result for the fitness distribution with increasing $N$ and fixed $K = 4$. The noise is Gaussian with $\sigma = 0.1$.

In Figure 2 we gave a plot of extinctions in an ecosystem of 10 000 species versus time. In this plot, there are brief periods of intense activity (mass extinctions) interspersed among periods of relative inactivity (background extinctions). To understand this behavior of the model, consider the fitness of the species in the system as a function of time. Starting with a well-adapted ecosystem in which most of the species have a high fitness (or tolerance to external influences) we allow the simulation to proceed. At each time step this process replaces $K$ species with new ones having randomly chosen barriers and fitnesses. These fitnesses have a reasonable chance of being below the original well-adapted val-ues, and hence are more likely to be wiped out by a small noise event. As described in Section II, the evolutionary process will proceed in coevolutionary avalanches, most of which will be small and involve few species. Bak and Sneppen (1993) have shown that there is a power law distribution for the size of these avalanches, with large ones occasionally occurring. When this happens, a large number of species have their fitnesses changed to new random values, and become more susceptible to external factors. As long as the noise level remains low, this has little effect. However, when a large noise event follows one of these large avalanches, a significant fraction of these species can be wiped out, giving rise to the large extinctions seen in the data. The important point here is that large extinctions do not necessarily occur only because of severe external stress, but also because of the coincidence of external stress and large coevolutionary avalanches, during which the susceptibility of species to external effects is enhanced. This is not to say that large extinctions can't occur due to environmental stresses alone, but that the effect of those stresses is enhanced by the variation in species' fitness brought about by the coevolutionary avalanches.

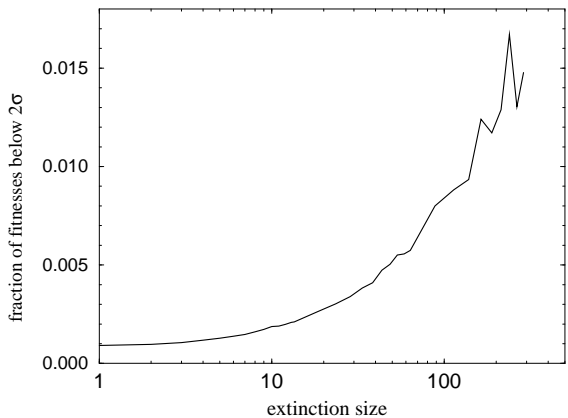

FIG. 8. The fraction of species having fitnesses below a certain threshold (twice the standard deviation of the noise in this case) immediately before extinctions of a certain size. Notice that the number of species having low fitness is higher immediately before a larger extinction, indicating that the large extinctions are the result of the coincidence of lower fitness with large environmental stresses. The data are for a system of size $N = 10\,000$ averaged over 1 million time steps with $\sigma = 0.2$.

There are several indicators in our simulation results that this is the correct explanation for the observed distribution of extinction sizes. For instance, we have calculated the fraction of species having a fitness below a certain threshold value immediately before an extinction. We then calculated the average of this result for each size of extinction. The results are shown in Figure 8, using a threshold of twice the standard deviation $\sigma$ of the Gaussian noise. This demonstrates that the fraction of species with low fitnesses is larger, on average, before a large extinction event. These events are therefore not just an effect of large environmental stresses.

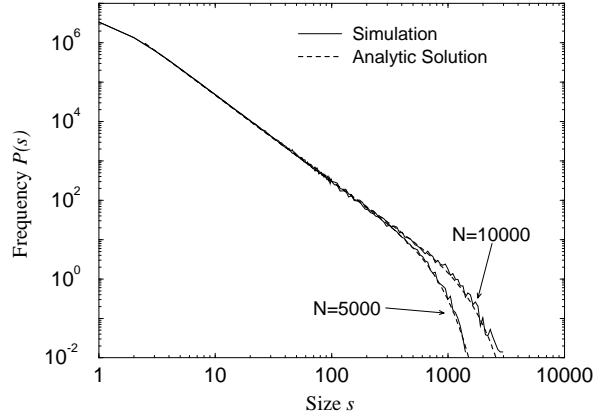

FIG. 9. Log–log plot of the distribution of extinction sizes in the model. The solid lines are simulation results, while the dashed lines are from the analytic solution, equation (24). The straight–line form of the graph indicates that the distribution is a power law, and the gradient of the line gives a value $2.02 \pm 0.03$ for the exponent of the power law. The parameters used here were $K = 4$, and $\sigma = 0.1$. We have averaged over 10 million time steps. Note the finite size cutoff discussed in the text.

Another important indicator is the distribution of extinction sizes. In Figure 9, we show logarithmic histograms of the size $s$ of extinctions versus their frequency of occurrence $P(s)$ for two different ecosystem sizes. The plot demonstrates that large extinctions are much less common than small ones. Furthermore, the distribution falls on a straight line for much of the graph, indicating that the distribution follows a power law:

$$P(s) \propto s^{-\alpha}, \qquad (26)$$

where from our plot we measure

$$\alpha = 2.02 \pm 0.03. \qquad (27)$$

(In our previous paper (Newman & Roberts, 1995), we published a slightly larger value for this exponent. The value given here represents the results of more extensive simulations and analysis.) In this plot finite–size effects are also apparent: the distribution $P(s)$ changes from a power law to a faster decay for large values of $s$, with the value of $s$ at which this change occurs increasing by a factor of 2 when we double the number of species in the system. This implies that the cutoff is an effect of the finite size of the system and not an intrinsic effect in the model.

Bak and Sneppen (1993) find a power–law distribution for the coevolutionary avalanches in their model, but with an exponent different from the one found here for extinctions. They suggest that avalanches and extinctions are in fact the same thing. We suggest that the avalanches are an important part of the mechanism giving rise to mass extinctions, but do not themselves represent extinctions. Because we require the coincidence of two unlikely events—severe environmental stress and a large coevolutionary avalanche—to produce a large extinction event, we expect our exponent $\alpha$ describing the distributions of extinctions to be larger than the one found by Bak and Sneppen for their avalanches. Having a large coevolutionary avalanche is not enough to lead to a mass extinction. If the environmental stresses are not high, then no large scale extinctions will occur. If environmental stresses are high, then large extinctions can occur, but they are enhanced by the existence of a large coevolutionary avalanche. Thus we expect large extinctions to be rarer than large avalanches, and the corresponding power law will be steeper. This is what we observe, since Bak and Sneppen measure a value of 1.35 for the exponent governing the avalanche distribution, which is considerably less than the 2.0 that we find for the extinctions.

The power–law form of the extinction distribution is a robust prediction of the model. The exponent $\alpha$ is also a fairly robust prediction of the model. We have simulated the model with various forms for the external noise, with different numbers of neighbors, and with several variations in the precise form of the dynamics, all without significantly changing the resulting value of $\alpha$ in equation (26). This illustrates that is not necessary to know the exact mechanism of interaction between species, or the precise form of external stresses that are responsible for mass extinctions or what their distribution is over time; our prediction for the power law is independent of such considerations. This is an important observation, since it should be possible to check this prediction against paleontological data such as that of Sepkoski (1993) or Benton (1995).

There are other features in the extinction data from our simulations that should also be visible in the fossil record. Consider, for example, the smaller extinctions preceding and following the largest ones, which we have called "precursors" and "aftershocks" (Newman & Roberts, 1995).

Precursors is the name we give to sets of smaller extinctions that precede a large extinction. Such precursors can be seen throughout the data from our simulations. In Figure 10, we see precursors leading to a large event in a portion of data drawn from the results in Figure 2. The explanation of this process is as follows. Large extinction events can occur when a large proportion of the population becomes less fit because the system is undergoing one or more coevolutionary avalanches. The next big noise event that comes along will wipe out a large portion of the species, provided the event occurs before the species evolve to fitter forms. However, there may be a long interval of time following an avalanche before a large external stress occurs in the system. In the meantime, the extinctions produced by small stresses will be enhanced because of the general unfitness of the species. Thus in the interval of time immediately preceding a major extinction event, we expect the background extinctions to be larger than normal. A similar effect is also seen in the fossil record. The mass extinction that occurred at the K–T boundary 65 million years ago was preceded by a period of about 3 million years during which many species on the planet were already dying out (Keller, 1989). The impact of a large meteor or comet at Chicxulub on the Yucatán Peninsula may have been the event that corresponds to the large noise event in our theory. However, this impact cannot explain why species were dying out before that event, and it has been suggested (see for example Kauffman (1993)) that this might be due to environmental stresses on a population that had become unfit for some other reason. This unfitness would also explain why the extinction effect of the impact was so great. In our model, we see the same effect, with the unfitness caused by large coevolutionary avalanches.

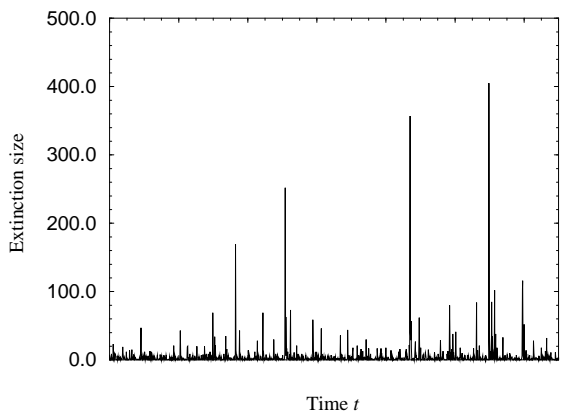

FIG. 10. An example of the 'precursor' effect described in Section IV. The time range here is 300 time steps. Compare this with Figure 2.

Another common pattern seen in the data from our simulations is the occurrence of a series of ex-tinctions of moderate size in the aftermath of a major extinction. These extinctions, which we call aftershocks, can be explained as follows. A large extinction wipes out many species, leaving empty many ecological niches. These niches are soon filled with new species. However, these species may not be well adapted to survive environmental stresses, since they have not evolved long enough to have felt the selective pressure of those stresses. Many of them will thus be wiped out by relatively small noise events that would not have affected more well–adapted populations. We expect to see moderate sized extinctions after large events in our model for exactly this reason. Figure 11 shows such a set of aftershocks, drawn from the extinction data in Figure 2. With the passage of time, the less fit species will be removed from the population and the general fitness will again increase to normal levels.

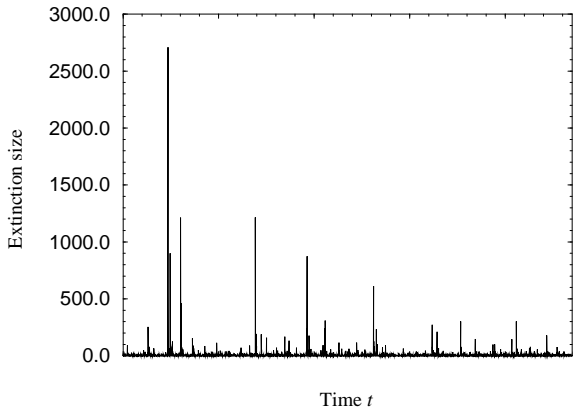

FIG. 11. An example of the 'aftershocks' described in Section IV. The time range in this plot is 700 time steps. Compare this with Figure 2.

Effects similar to our aftershocks are seen in the fossil record. The increase in speciation following mass extinctions is a common pattern (Sepkoski, 1993). For example, during the Cambrian explosion of 570 million years ago, a large number of species appeared over a short geological time span and colonized a variety of ecological niches. Many of these species disappeared soon afterwards, possibly because they were not able to cope with the ongoing changes in their environment. Thus extinction, as well as speciation, appears to have been at a maximum during this period.

As an aside, we note that abrupt extinctions in the fossil record could be smeared out due to processes affecting the deposition and preservation of fossils (taphonomic processes). Sampling can give the appearance of smoother extinctions. This is called the Signor–Lipps effect (Signor & Lipps, 1982). Such processes could lead to the belief that a particularly large extinction is instead a series of small, though still significant, ones. A series of smaller events could also be coalesced into one due to erosion or nondeposition of sediments during the critical time interval, or due to the coarseness of the geological time scale. In some sense this latter effect is also present in our model, since we have not determined the fundamental time scale in the model. If we were to look at the extinctions in the model on a coarser time scale, we would see larger extinctions that are really the combination of underlying smaller extinctions.

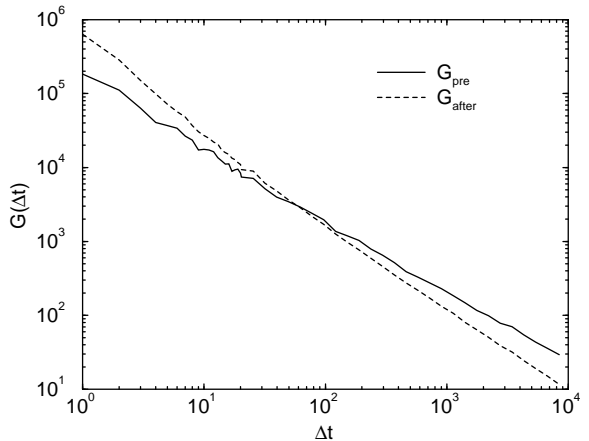

FIG. 12. Log–log plot of correlations between extinctions, $G_{\rm pre}(\Delta t)$ for precursor extinctions (solid line) and $G_{\rm after}(\Delta t)$ for aftershocks (dashed line). Note that the power–law decays for the two correlation functions are not the same.

In order to quantify the notion of aftershocks and precursors, we define correlation functions for the extinction sizes $s(t)$ as

$$G_{\rm pre}(\Delta t) = \langle s(t-\Delta t)\,s(t) \rangle_{s(t-\Delta t)<s(t)},$$

$$G_{\rm after}(\Delta t) = \langle s(t)\,s(t+\Delta t) \rangle_{s(t)>s(t+\Delta t)}. \qquad (28)$$

The precursors are described by $G_{\rm pre}(\Delta t)$ and the aftershocks by $G_{\rm after}(\Delta t)$. The brackets $\langle \ldots \rangle$ represent averages over time. In Figure 12 we show results for these correlation functions taken from our simulations. The slope of the lines indicates that the largest extinction events tend to be correlated with smaller events immediately before and after. It also appears from the figure that the correlation functions decay as a power law in $\Delta t$ for both the precursors and aftershocks:

$$\begin{aligned} G_{\rm pre}(\Delta t) &= (\Delta t)^{-\beta_-}, \\ G_{\rm after}(\Delta t) &= (\Delta t)^{-\beta_+}. \end{aligned} \qquad (29)$$

We find from the numerical simulations that $\beta_- = 0.95 \pm 0.02$ and $\beta_+ = 1.10 \pm 0.02$.

It has been suggested that the extinctions seen in the fossil record have a periodicity of 26 million years (Raup & Sepkoski, 1984, 1986; Sepkoski, 1990). We have tried simulating this effect by periodically varying the strength of the external noise in our model. We have used a noise function of the form

$$r_{\text{periodic}}(t) = r(t) + \sigma(1 + \sin(\omega t)), \qquad (30)$$

where $r(t)$ is the noise function used in the earlier simulations (which had a Gaussian distribution of width $\sigma$), and $\omega$ is the frequency of the periodicity. In our simulations we chose a value of $\omega = \frac{2\pi}{200}$. Data from one of these simulations are shown in Figure 13. The system appears approximately periodic by visual inspection, and we can test this conclusion by examining the power spectrum of the time series. In Figure 14 we show this power spectrum, which is strongly peaked at the appropriate frequency, indicating that the periodic external factors do indeed cause the system to have periodic extinctions, though we can still see moderate–sized extinctions occurring when the noise is not at a maximum. We can also see clusters of moderate–sized extinctions occurring around the points where the noise peaks in time. These clusters of extinctions are similar in nature to the precursors and aftershocks discussed above.

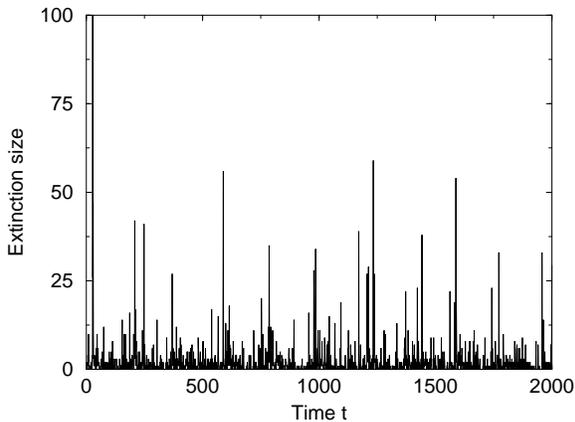

FIG. 13. Time series for periodic external stresses. Here we have used $K = 4$, $N = 512$, and $\sigma = 0.2$. The external "noise" is of the form $r_{\text{periodic}}(t) = r(t) + \sigma(1 + \sin(\omega t))$ where $r(t)$ is the normal nonperiodic noise function and the frequency $\omega$ was chosen to be $\frac{2\pi}{200}$.

We have also looked at the distribution of extinction sizes in the periodic case. The statistics for these simulations were not as good as in the non-periodic case (we had a smaller ecosystem $N = 512$ and a shorter run of 131 072 time steps), but we still see power–law behavior in the distribution, and we extract a value of $\alpha = 2.3 \pm 0.2$ for the exponent. This is similar to the result found above for the non-periodic version of the model, further demonstrating that the dynamics of the model results in power laws regardless of the particular form of the external influences.

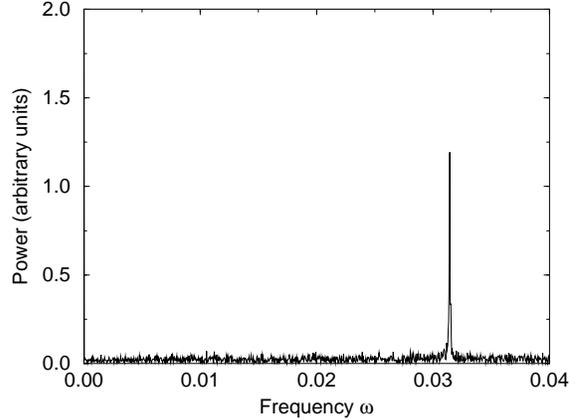

FIG. 14. Power spectrum of the time series with periodic external influences. Note the peak at $\omega = \frac{2\pi}{200}$.

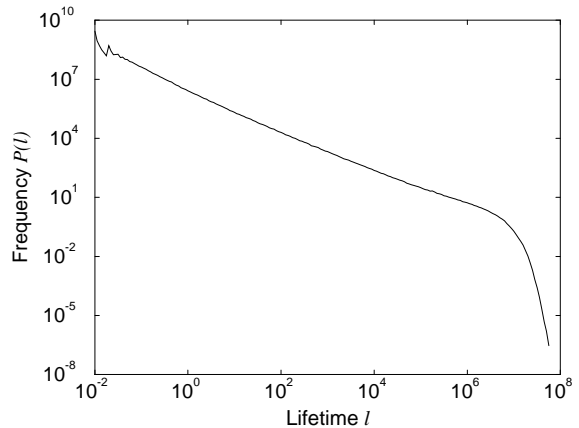

FIG. 15. A log-log histogram of the distribution of species lifetimes measured from when a species first appears to when it becomes extinct, for a simulation lasting ten million time-steps. The straight-line form of the histogram indicates a power–law falloff in the number of species surviving to longer and longer times. The exponential tail at very large lifetimes is due to the finite size of the simulated system. Similar behavior is seen in the fossil data of Sepkoski (see Raup (1991) page 55) at the genera level. This simulation was performed with $N = 10\,000$, $K = 4$, and $\sigma = 0.2$.

We can also measure within our model a distribution of species lifetimes—see Figure 15. Note that the plot is a log–log plot, and the straight line form of the plot demonstrates that this distribution also decays as a power law, $P(l) \sim l^{-\gamma}$. We measure a value of $\gamma = 0.99 \pm 0.03$ for the exponent. For large lifetimes, the distribution falls off exponentially. This is a result of the finite size of the systems we have studied. It is unlikely for a species with a high fitness to become extinct simply because of the external noise. However, since $K - 1$ randomly chosen species become extinct at each step in the simulation by virtue of the evolution of a neighboring species, the chances of any one species surviving become exponentially small for times much longer than approximately $N/(K-1)$ time steps.

## V. CONCLUSIONS

We have considered a mechanism for evolution and extinction where large extinctions arise as a result of the coincidence of coevolutionary avalanches which lower the general fitness of the species in an ecosystem, and large environmental stresses which wipe out the less fit species. We have developed a simple mathematical model which describes such a process and gives predictions about the distributions of extinction sizes and species lifetimes. The extinctions appear to have a power-law distribution with an exponent independent of the form of the external stresses. The prediction of power-law decay should be testable against the fossil record.

There are a number of extensions that could be added to the model, which we have mentioned at various points in the text. These include, but are not limited to, true speciation events, population size effects, different interactions between species, more sophisticated spatial models, and environmental stress effects that vary over spatial regions. These might all be interesting to add, but the model would quickly become cumbersome, and it would be difficult to sort out causes and effects. We have confined ourselves to the simplest possible model that we feel captures the essential biological interactions.

## VI. ACKNOWLEDGMENTS

We would like to thank A. L. Killen, D. M. Raup, H. K. Reeve, J. P. Sethna, K. Sneppen, and R. D. Yanai for useful discussions and comments. This work was supported in part by the NSF through grant numbers DMR–91–18065 and ASC–94–04936 and by the Cornell Theory Center and the Hertz Foundation. The simulations were performed using computing facilities at the Cornell Theory Center.